\RequirePackage{amsmath} 
\documentclass[runningheads]{llncs}
\usepackage{hyperref}
\hypersetup{
    colorlinks,
    linkcolor={red!50!black},
    citecolor={blue!50!black},
    urlcolor={blue!80!black}
}
\usepackage{orcidlink}
\usepackage[T1]{fontenc}
\usepackage{mathtools}
\usepackage{amsfonts,amssymb}
\usepackage[misc,geometry]{ifsym}
\usepackage{xfrac}
\usepackage{pifont}
\usepackage{graphicx}
\usepackage{multirow}
\usepackage{xcolor}
\usepackage{tikz}
\usepackage[ruled,vlined,linesnumbered,nofillcomment]{algorithm2e}
\usepackage[caption=false]{subfig}
\usepackage{float}
\usepackage{lmodern}
\usepackage[textsize=scriptsize]{todonotes}
\usepackage{cleveref}

\newcommand{\arxiv}[2]{#1} 

\DeclareMathSymbol{\smin}{\mathbin}{AMSa}{"39}
\newcommand{\comp}{\text{comp}}
\newcommand{\transpose}{\intercal}
\newcommand{\tensor}{\otimes}
\newcommand{\neglit}[1]{\overline{#1}}

\SetFuncArgSty{textnormal}
\SetArgSty{textnormal}
\SetKwProg{Fn}{def}{}{}
\SetFuncSty{textsc}
\SetVlineSkip{0pt}
\DontPrintSemicolon

\SetKwComment{Comment}{$\triangleright$ }{}
\SetCommentSty{mycommfont}
\SetKw{Or}{\textnormal{\textbf{or}}}
\SetKw{And}{\textnormal{\textbf{and}}}

\usetikzlibrary{patterns,shapes,arrows,fit,calc,positioning,automata}
\tikzset{main/.style={draw,circle}} 
\tikzset{leaf/.style={draw,minimum width=1.2em,minimum height=1.2em}} 
\tikzset{e0/.style={draw,->,dotted,>=latex}}
\tikzset{e1/.style={draw,->,>=latex}}

\newcommand{\follow}[2]{#2[#1]}
\newcommand{\makenode}{\textsc{MakeNode}}
\newcommand{\makeleaf}{\textsc{MakeLeaf}}
\newcommand{\ddvar}[1]{\textsf{var}(#1)}
\newcommand{\ddval}[1]{\textsf{val}(#1)}

\crefname{section}{Sec.}{Secs.}
\Crefname{section}{Section}{Sections}
\crefname{line}{l.}{ll.}
\Crefname{line}{line}{lines}
\crefname{equation}{Eq.}{Eqs.}
\Crefname{equation}{Equation}{Equations}
\crefname{algorithm}{Alg.}{Algs.}
\Crefname{algorithm}{Algorithm}{Algorithms}
\crefname{theorem}{Thm.}{Theorems}
\Crefname{theorem}{Theorem}{Theorems}
\crefname{corollary}{Cor.}{Corollaries}
\Crefname{corollary}{Corollary}{Corollaries}
\crefname{example}{Ex.}{Examples}
\Crefname{example}{Example}{Examples}

\RequirePackage{graphicx}
\usepackage[firstpageonly=true,angle=0,vpos=.147\paperheight,hpos=.393\linewidth,vanchor=t,hanchor=l]{draftwatermark}
\SetWatermarkText{%
\raisebox{-3.1cm}
{\begin{minipage}{\linewidth}\centering\href{https://doi.org/10.5281/zenodo.15322337}{\includegraphics[width=12.5mm]{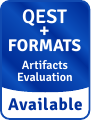}}\hfill\includegraphics[width=12.5mm]{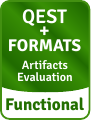}\end{minipage}}%
}

\begin{document}

\title{Numerical Errors in Quantitative System Analysis With Decision Diagrams}
\titlerunning{Numerical Errors in Quantitative System Analysis With Decision Diagrams}
%
\author{Sebastiaan Brand$^{\text{(\Letter})}$\orcidlink{0000-0002-7666-2794} \and
Arend-Jan Quist\orcidlink{0000-0002-6501-2112} \and \\
Richard M.K. van Dijk\orcidlink{0000-0002-5796-7883} \and
Alfons Laarman\orcidlink{0000-0002-2433-4174}}
\authorrunning{S. Brand et al.}
%
\institute{Leiden Institute of Advanced Computer Science, Leiden University, The Netherlands
\email{s.o.brand@liacs.leidenuniv.nl}}

\maketitle

\begin{abstract}
Decision diagrams (DDs) are a powerful data structure that is used to tackle the state-space explosion problem, not only for discrete systems, but for probabilistic and quantum systems as well.
While many of the DDs used in the probabilistic and quantum domains make use of floating-point numbers, this is not without challenges. Floating-point computations are subject to small rounding errors, which can affect both the correctness of the result and the effectiveness of the DD's compression.
In this paper, we investigate the numerical stability, i.e. the robustness of an algorithm to small numerical errors, of matrix-vector multiplication with multi-terminal binary decision diagrams (MTBDDs). Matrix-vector multiplication is of particular interest because it is the function that computes successor states for both probabilistic and quantum systems.
We prove that the MTBDD matrix-vector multiplication algorithm can be made numerically stable under certain conditions, although in many practical implementations of MTBDDs these conditions are not met.
Additionally, we provide a case study of the numerical errors in the simulation of quantum circuits, which shows that the extent of numerical errors in practice varies greatly between instances.
\end{abstract}

\section{Introduction}
The analysis and verification of discrete, probabilistic, or quantum systems involves exploring a space of possible behaviors that is exponential in the system's description. Decision diagrams (DDs) are a data structure that can be used to symbolically represent huge subspaces of such systems~\cite{mcmillan1992symbolic,hensel2022storm,viamontes2004high,miller2006qmdd}, lifting the computation to this representation. 

More specifically, DDs are directed graphs that represent pseudo-Boolean functions of $n$ variables with a co-domain of real or complex numbers in the probabilistic and quantum settings.
Every node in a DD represents a sub-function, and DDs achieve compactness by merging nodes that represent equivalent sub-functions.
While it is possible to represent values in the co-domain exactly, using for example rational~\cite{hensel2022storm} or other algebraic representations~\cite{niemann2020overcoming,quist2025exact}, this is often not practical.
In the probabilistic domain, the commonly-used value iteration algorithm becomes inefficient when using rational numbers~\cite{hensel2022storm}, while in the quantum domain, algorithms such as the ubiquitous quantum Fourier transform cannot be represented exactly with algebraic representations such as presented in~\cite{niemann2020overcoming}.

Many DD packages therefore include floating-point representation for their values~\cite{sanner2005affine,miller2006qmdd,somenzi2015cudd,vandijk2017sylvan,zulehner2018advanced}.
However, floating-point representations come with their issues: they are often not exact, and operations on them can introduce small rounding errors.
These errors not only affect the correctness of the output but can also prevent equivalent sub-functions from being recognized, thus undermining the DD's compression.
Issues with floating-point errors have been reported for various types of DDs~\cite{sanner2005affine,zulehner2019efficiently,niemann2020overcoming,sistla2023symbolic}. 
In the implementation of many DDs that make use of floating-point values, authors mitigate this problem by introducing a small merging threshold~\cite{sanner2005affine,sanner2010approximate,somenzi2015cudd,zulehner2019efficiently}, which we will call $\delta$. Two floating-point values $a$ and $a'$ are considered equivalent if $|a - a'| \leq \delta$.
However, as the framework we introduce will show, this value merging is now a new source of potential errors, which we will call ``merging errors''.

To gain a better insight into the effects of floating-point errors that occur during the analysis of probabilistic and quantum systems using DDs, we perform error analysis on the matrix-vector multiplication algorithm for multi-terminal binary decision diagrams (MTBDDs). We pick matrix-vector multiplication specifically because it is ubiquitous in the verification and analysis of discrete, probabilistic, and quantum systems, where it is used to exhaustively explore the system's behavior by computing successor states under the system's transition relation~\cite{mcmillan1992symbolic,nielsen2010quantum,katoen2016probabilistic}.

When analyzing how numerical errors grow during the analysis of such systems, a metric for the size of the system is needed. Two options present themselves here. We can either consider the size of the original system description ---for instance, a Bayesian network or a quantum circuit--- or the size of a symbolic representation of the system's behavior ---such as the states and transition relations recorded in a decision diagram.
The latter is important in practice because it is often used in efficient symbolic analysis methods~\cite{mcmillan1992symbolic,sang2005performing,zulehner2018advanced}. 
In the context of the analysis of combinatorial systems, previous studies on the numerical stability of linear algebra effectively consider this measure, and the relevant vectors and matrices are of exponential length in the size of the original system.
The former, however, represents the more strict viewpoint, as the system's description can be exponentially more succinct than the DDs computed when exhaustively exploring its behavior. 
This viewpoint is therefore considered in the complexity analysis of such symbolic methods~\cite{feigenbaum1998complexity}.
Additionally, in practice, it is arguably more relevant to understand how the error grows with the size of the system under analysis. For example, in the numerical analysis of a quantum system, it is interesting to understand how the numerical errors grow with the number of quantum bits that make up the system, or in probabilistic inference, we might consider the size of the Bayesian network.
\emph{For this reason, we study numerical stability from the stringent perspective of the original system description.}

We obtain bounds on the worst-case errors, stated in \Cref{thm:main_thm} for arbitrary matrices and vectors, and in \Cref{thm:main_thm-prob-quantum} for probabilistic and quantum systems specifically. 
The bounds are parameterized in three variables: the size of the underlying system, expressed in the number of Boolean variables $n$ used to represent its states, the maximum rounding error $\varepsilon$ which depends on the floating-point precision, and the merging threshold $\delta$ which allows values that are $\delta$-close to be merged.

In line with error analyses of other linear algebra algorithms~\cite{demmel2007fastmatrix,demmel2007fastlinear}, we analyze the \emph{forward error}, which assumes the input to be exact and measures how floating-point errors accumulate in the output. Additionally, we consider \emph{componentwise errors}, i.e. errors on the individual elements of the output vector. We consider the algorithm \emph{forward stable} if the forward error grows at most polynomially with the size $n$ of the underlying system.

Our theoretical findings can be summarized as follows.
\begin{enumerate}
    \item The MTBDD matrix-vector multiplication algorithm is not componentwise forward stable in $n$ for arbitrary vectors and matrices.
    \item The MTBDD matrix-vector multiplication algorithm is componentwise forward stable in $n$ for probabilistic and quantum systems when the merging threshold $\delta \in O(2^{-n})$, or simply $\delta = 0$.
    \item The MTBDD matrix-vector multiplication algorithm is not componentwise forward stable in $n$ for quantum or probabilistic systems if $\delta$ is constant and non-zero.
\end{enumerate}

Finally, we also provide a case study of numerical errors when using MTBDDs to simulate quantum circuits, where we evaluate the effect of the choice of $\delta$ on both the DD size as well as the errors on the output. For this evaluation, we make use of an MTBDD implementation that allows for arbitrary-precision floating-point values.\footnote{This implementation is part of the Q-Sylvan DD package and can be found online here: \url{https://github.com/System-Verification-Lab/Q-Sylvan}}
This case study shows that, in practice, the effects of numerical errors can greatly vary between different types of instances.


\section{Preliminaries}
\label{sec:prelims}

In this section, we briefly explain MTBDDs, how floating-point errors can cause issues with a DD's compactness, and the basics of floating-point error analysis.

\subsection{MTBDDs}
MTBDDs, also referred to as algebraic decision diagrams (ADDs)~\cite{frohm1993algebraic,fujita1997multi}, are rooted, directed, acyclic graphs, and can be used to represent pseudo-Boolean functions $f : \{0, 1\}^n \to \mathbb{D}$, where the domain $\mathbb{D}$ can for example be real or complex numbers. The function $f(x)$ can also be seen as a $2^n$-dimensional vector, where $x$ specifies an index in the vector and $f(x)$ gives the corresponding value.
MTBDDs have two types of nodes: leaf nodes, which have a value $\ddval{\cdot} \in \mathbb{D}$, and internal nodes, which each have a variable $\ddvar{\cdot} \in \{x_0, \dots x_{n-1}\}$. MTBDDs are ordered; that is, on every path from the root to a leaf, the variables are encountered in the same order, $x_{j} \prec x_{j+1}$, although variables may be skipped.

An internal MTBDD node $v$ can be understood as the Shannon decomposition of the function $f^v$ it encodes:
\begin{align*}
   \begin{tikzpicture}[node distance=10mm, inner sep=1.5pt, auto, thick] 
    	\node[] (0) [] {};
    	\node[main,label=right:{$v$}] (1) [node distance=7mm,below of=0] {$x_j$};
    	\node[] (2) [below left of=1] {$v[0]$};
    	\node[] (3) [below right of=1] {$v[1]$};
    	\draw[->] (0) to node [] {$f^v$} (1) ;
    	\draw[->,dashed] (1) to (2);
    	\draw[->] (1) to (3);
        \node[] (eq) [right = 2cm of 1] {$f^v = \neglit x_j \cdot f^v_{|0} + x_j \cdot f^v_{|1}$};
    \end{tikzpicture}
\end{align*}
This decomposition can be read as ``if $x_j = 0$ then $\follow{0}{v}$, else $\follow{1}{v}$'', where $\follow{0}{v}$ and $\follow{1}{v}$ correspond to the sub-functions of $f^v$ where $x_j$ has been set to 0 and 1 respectively.
In general we let $f_{|0}$ ($f_{|1}$) denote the sub-function of $f$ where the first variable of $f$ is set to 0 (1).
Every path through the MTBDD corresponds to a single value of the vector/function it represents.
For example, the value $f(x_0,x_1,x_2)_{|100} = f(100) = 3$ can be retrieved from the MTBDD in \Cref{fig:mtbdd} by following the 1 (solid) edge for $x_0$, and the 0 (dashed) edge for $x_1$ and $x_2$.

\begin{figure}[b]
    \centering
    \subfloat[without $\delta$-merging]{%
        \scalebox{1}{\begin{tikzpicture}[auto, thick,node distance=1.cm,inner sep=1.5pt]
    \newcommand\dist{.34cm}
    \newcommand\sep{.7cm}
    \newcommand\sepp{0.15cm}
    \newcommand\seppp{-.15cm}
    \node[] (0) [] {};
    \node[main] (x0)  [node distance=.7cm,below of=0] {$x_0$};
    \node[main] (x1)  [below right = \dist and \sep of x0] {$x_1$};
    \node[main] (x21) [below left  = 3.1*\dist and \sepp of x0] {$x_2$};
    \node[main] (x22) [below left  = \dist and \sepp of x1] {$x_2$};
    \node[main] (x23) [below right = \dist and \sepp of x1] {$x_2$};
    \node[leaf] (l1)  [below left  = 1.1*\dist and \seppp of x21] {$\smin2$};
    \node[leaf] (l2)  [below left  = 1.1*\dist and \seppp of x22] {$1$};
    \node[leaf] (l3)  [below right = 1.1*\dist and \seppp of x22] {$3$};
    \node[leaf] (l4)  [below left  = 1.1*\dist and \seppp of x23] {$2$};
    \node[leaf] (l5)  [below right = 1.1*\dist and \seppp of x23] {$2.0...04$};
    \draw[e1] (0) to (x0);
    \draw[e0,bend right=20] (x0) to (x21);
    \draw[e1,bend left=10] (x0) to (x1);
    \draw[e1] (x21) to (l1);
    \draw[e0] (x21) to (l2);
    \draw[e0] (x1) to (x22);
    \draw[e1] (x1) to (x23);
    \draw[e1] (x22) to (l2);
    \draw[e0] (x22) to (l3);
    \draw[e0] (x23) to (l4);
    \draw[e1] (x23) to (l5);
\end{tikzpicture}}%
        \label{fig:mtbdd}%
    }\hspace{1cm}
    \subfloat[with $\delta$-merging]{%
        \scalebox{1}{\begin{tikzpicture}[auto, thick,node distance=1.cm,inner sep=1.5pt]
    \newcommand\dist{.34cm}
    \newcommand\sep{.7cm}
    \newcommand\sepp{0.15cm}
    \newcommand\seppp{-.15cm}
    \node[] (0) [] {};
    \node[main] (x0)  [node distance=.7cm,below of=0] {$x_0$};
    \node[main] (x1)  [below right = \dist and \sep of x0] {$x_1$};
    \node[main] (x21) [below left  = 3.1*\dist and \sepp of x0] {$x_2$};
    \node[main] (x22) [below left  = \dist and \sepp of x1] {$x_2$};
    \node[leaf] (l1)  [below left  = 1.1*\dist and \seppp of x21] {$\smin2$};
    \node[leaf] (l2)  [below left  = 1.1*\dist and \seppp of x22] {$1$};
    \node[leaf] (l3)  [below right = 1.1*\dist and \seppp of x22] {$3$};
    \node[leaf] (l4)  [below left  = 1.1*\dist and \seppp of x23] {$2$};
    \draw[e1] (0) to (x0);
    \draw[e0,bend right=20] (x0) to (x21);
    \draw[e1,bend left=10] (x0) to (x1);
    \draw[e1] (x21) to (l1);
    \draw[e0] (x21) to (l2);
    \draw[e0] (x1) to (x22);
    \draw[e1] (x22) to (l2);
    \draw[e0] (x22) to (l3);
    \draw[e1] (x1) to (l4);
\end{tikzpicture}}%
        \label{fig:mtbdd-merged}%
    }
    \caption{An MTBDD that encodes the vector $\begin{pmatrix} 1 & 1 & \smin 2 & \smin 2 & 3 & 1 & 2 & \hat 2\end{pmatrix}^\intercal$, with $\hat 2 = 2.0...04$ in \protect\subref{fig:mtbdd} and $\hat 2 = 2$ in \protect\subref{fig:mtbdd-merged}.}
    \label{fig:mtbdd-example}
\end{figure}

Decision diagram algorithms are typically defined recursively, using the Shannon decomposition described above.
As an example, the addition of two vectors $\vec a$ and $\vec b$ can be split up recursively as
{%
\setlength\arraycolsep{3pt}%
$
\vec a + \vec b = 
    \begin{pmatrix}
        \vec a_{|0} & \vec a_{|1}
    \end{pmatrix}^\transpose +
    \begin{pmatrix}
        \vec b_{|0} & \vec b_{|1}
    \end{pmatrix}^\transpose =
    \begin{pmatrix}
        \vec a_{|0} + \vec b_{|0} & \vec a_{|1} + \vec b_{|1}
    \end{pmatrix}^\transpose =
    \begin{pmatrix}
        \vec c_{|0} & \vec c_{|1}
    \end{pmatrix}^\transpose =
    \vec c
$.
}%
The corresponding MTBDD algorithm is given in \Cref{alg:mtbdd-plus}.
Caching of previously computed values is used to avoid redundant recursive calls. If $A$ and $B$ are leaves, a leaf with the sum of their values is returned. The \makeleaf{} function ensures that leaves are canonical, i.e. for every unique value there is only one corresponding leaf node. 
If a leaf with value $\ddval{A} + \ddval{B}$ already exists, that leaf is returned. Otherwise, a new leaf is created. 
The \makenode{} function does the same for internal 
nodes. This can be done efficiently since the DD is built from the bottom up, thus allowing \mbox{\makenode{}} to rely on the already canonical representation of child nodes.

Similarly to vector addition, matrix-vector multiplication can be split up as
\begin{align*}
    M \vec v =
    \begin{pmatrix}
        M_{|00} & M_{|01} \\
        M_{|10} & M_{|11}
    \end{pmatrix}
    \begin{pmatrix}
        \vec v_{|0} \\ \vec v_{|1}
    \end{pmatrix} = 
    \begin{pmatrix}
        M_{|00} \vec v_{|0} + M_{|01} \vec v_{|1} \\
        M_{|10} \vec v_{|0} + M_{|11} \vec v_{|1}
    \end{pmatrix}.
\end{align*}
The corresponding pseudo-code is given in \Cref{alg:mtbdd-mult}. To efficiently access the submatrices $M_{|i,j}$, variables representing column and row indices are interleaved.

\begin{algorithm}[t]
\SetKwFunction{add}{Plus}
\Fn{\add{MTBDD $A$, MTBDD $B$}}{
    \vspace{-1.25em}\Comment*[r]{assuming $\ddvar{A} = \ddvar{B}$}

    \BlankLine
    \If{$A$ and $B$ are leaves}{
        \vspace{-1.25em}\Comment*[r]{terminal case}
        \Return \makeleaf($\ddval{A} + \ddval{B}$)
    }

    \BlankLine\BlankLine
    \lIf{exists ($R \gets$ \textnormal{cache[\add,$A,B$]})}{
        \Return{$R$}
    }
    
    \BlankLine\BlankLine
    
    $R_0 \gets$ \add{$\follow{0}{A}$, $\follow{0}{B}$} \;
    $R_1 \gets$ \add{$\follow{1}{A}$, $\follow{1}{B}$} \;
    $R \gets \makenode(\ddvar{A}, R_0, R_1)$ \;

    \BlankLine\BlankLine
    \textnormal{cache[\add,$A,B$]} $\gets R$\; 
    \BlankLine\BlankLine
    
    \Return{$R$}
    \scalebox{.95}{\begin{tikzpicture}[node distance=10mm, inner sep=1.5pt, auto, thick, overlay, xshift=4.5cm, yshift=1.5cm] 
	\node[] (0a) [] {};
	\node[main] (1a) [node distance=7mm,below of=0a] {$x_i$};
	\node[] (2a) [below left  = 4mm and -2mm of 1a] {$\follow{0}{A}$};
	\node[] (3a) [below right = 4mm and -2mm of 1a] {$\follow{1}{A}$};
	\draw[->] (0a) to node [left,pos=.3] {$A$} (1a) ;
	\draw[->,dashed] (1a) to (2a);
	\draw[->] (1a) to (3a);

	\node[] (plus) [right = .3cm of 1a] {$+$};

	\node[main] (1b) [right = .3cm of plus] {$x_i$};
	\node[] (0b) [node distance=7mm, above of=1b] {};
	\node[] (2b) [below left  = 4mm and -2mm of 1b] {$\follow{0}{B}$};
	\node[] (3b) [below right = 4mm and -2mm of 1b] {$\follow{1}{B}$};
	\draw[->] (0b) to node [left,pos=.3] {$B$} (1b) ;
	\draw[->,dashed] (1b) to (2b);
	\draw[->] (1b) to (3b);

	\node[] (equals) [right = .6cm of 1b] {$=$};

	\node[main] (1c) [right = 1.0cm of equals] {$x_i$};
	\node[] (0c) [node distance=7mm, above of=1c] {};
	\node[] (2c) [below left  = 4mm and -1.7mm of 1c] {$\follow{0}{A}$+$\follow{0}{B}$};
	\node[] (3c) [below right = 4mm and -1.7mm of 1c] {$\follow{1}{A}$+$\follow{1}{B}$};
	\draw[->] (0c) to node [left,pos=.3] {$R$} (1c) ;
	\draw[->,dashed] (1c) to (2c);
	\draw[->] (1c) to (3c);
\end{tikzpicture}}
}

\caption{Vector addition using MTBDDs.}
\label{alg:mtbdd-plus}
\end{algorithm}

\begin{algorithm}[t]
\SetKwFunction{mult}{Multiply}
\Fn{\mult{MTBDD $M$, MTBDD $V$}}{
    \Comment*[l]{$M$: MTBDD over $2n$ variables representing a $2^n \times 2^n$ matrix}
    \Comment*[l]{$V$: MTBDD over $n$ variables representing a $2^n$-sized vector}
    \BlankLine
    \If{$M$ and $V$ are leaves}{
        \Return \makeleaf($\ddval{M} \cdot \ddval{V}$)
    }

    \BlankLine\BlankLine
    \lIf{exists ($R \gets$ \textnormal{cache[\mult,$M,V$]})}{
        \Return $R$
    }

    \BlankLine\BlankLine
    $R_{00} \gets $\mult{$\follow{00}{M},\follow{0}{V}$} \;
    $R_{10} \gets $\mult{$\follow{10}{M},\follow{0}{V}$} \;
    $R_{01} \gets $\mult{$\follow{01}{M},\follow{1}{V}$} \;
    $R_{11} \gets $\mult{$\follow{11}{M},\follow{1}{V}$} \;
    \BlankLine\BlankLine
    $R_0 \gets $\add{$R_{00},R_{10}$} \;
    $R_1 \gets $\add{$R_{01},R_{11}$} \;
    \BlankLine\BlankLine
    $R \gets \makenode(\ddvar{v},R_0, R_1)$

    \BlankLine\BlankLine
    \textnormal{cache[\mult,$M,V$]} $\gets R$\; 
    \BlankLine\BlankLine

    \Return $R$
}

\caption{Matrix-vector multiplication using MTBDDs.}
\label{alg:mtbdd-mult}
\end{algorithm}

\subsection{Floating-Point Values in DDs}
\label{sec:prelims-floats-in-dds}
Floating-point operations are typically not exact. For example, $0.1 + 0.2 - 0.1$ might yield 0.20000000000000004 instead of 0.2. For an MTBDD that has floating-point values in its leaves this creates a problem: new leaves are introduced that prevent the DD from staying as compact as it might have been with exact computations.
To allow nodes to merge despite possible floating-point errors, the merging rule for leaves might be slightly relaxed. Specifically, we can allow two leaf nodes $u$ and $v$ to merge when $|\ddval{u} - \ddval{v}| \leq \delta$, for some choice of $\delta$. An example is given in \Cref{fig:mtbdd,fig:mtbdd-merged}.
The allowance of approximate equality between floating-point values is used in implementations of various decision diagrams~\cite{sanner2005affine,sanner2010approximate,somenzi2015cudd,zulehner2019efficiently}.

The actual ``merging'' of floating-point values can be handled in different ways. Typically, \makeleaf{$(a)$} returns a value $a'$ that satisfies $|a' - a| \leq \delta$, if there is such an $a'$ among existing leaves. The value $a$ is then completely forgotten. When multiple values are $\delta$-close to $a$, for example one slightly greater and one slightly smaller than $a$, which value is returned depends on the specific DD implementation.
To keep our analysis mostly agnostic of such implementation details, the only assumption we make is that \makeleaf{$(a)$} returns a leaf node with value $a'$, such that $|a - a'| \leq \delta$.

While floating-point rounding errors are multiplicative (i.e. the error is relative to the magnitude of the value), errors introduced by $\delta$-merging are additive. These additive errors have a significantly greater impact than multiplicative errors, as illustrated in \Cref{tab:errors-for-concrete-parameters}. Despite this, having a relative merging rule (e.g. merge $a$ and $a'$ when $|\frac{a}{a'} - 1| \leq \delta$) is not desirable either, since this will prevent values from merging with $0$, taking away a potentially larger source of DD space reduction.

\subsection{Error Analysis and Numerical Stability}
\label{sec:prelims-error-analysis}
The analysis of floating-point errors in algorithms is a well-established research domain. In this section, we recap the terms and definitions that are relevant to this work, for which we take~\cite{higham2002accuracy} as a basis.

The general assumption for analyzing the error of floating-point operations is that every operation $\textsc{op}(\cdot,\cdot)\in\{+,-,\times\}$ on two real or complex\footnote{Rounding errors for complex arithmetic can be slightly greater than for real numbers since complex multiplication requires multiple operations. However, this only affects the value of $\varepsilon$ ($\varepsilon_{\text{complex}} \approx 2\varepsilon_{\text{real}}$) and does not affect the analysis.} numbers outputs the correct value up to a rounding error, i.e. the computed value $\textsc{op}(a,b)_{\comp}$ is equal to $\textsc{op}(a,b)(1+\theta)$, for some $\theta$ with $|\theta|\leq\varepsilon$. Here, $\varepsilon$ depends on the numerical precision of floating-point numbers~\cite[Eq.(2.4)]{higham2002accuracy} and can be assumed to scale as $\varepsilon \approx 2^{-b}$, where $b$ is the number of bits representing the mantissa~\cite{demmel2007fastmatrix}.
The maximum rounding error $\varepsilon$ is typically of the order $2^{-53} \approx 10^{-16}$ ($2^{-24} \approx 10^{-7}$) for 64-bit (32-bit) floating-point arithmetic.

For a computation $f$, where $f$ describes a concrete algorithm, on an input $x$ we denote $\hat y = f_\comp(x)$
as the computed output, and $y = f(x)$ as the exact output.
The quality of a computed $\hat y$ can be measured by either its forward or backward error.
The forward error is the difference $|y - \hat y|$ between the computed value and the exact value, while the backward error is the smallest $|\Delta x|$ such that $f(x + \Delta x) = \hat y$.
An algorithm is called forward (backward) stable if the forward (backward) error is small, where the definition of ``small'' is context dependent.
Alternatively put, an algorithm is forward stable if $\hat y$ is only slightly wrong for the correct input, and backward stable if $\hat y$ is the correct output for a slightly wrong (or perturbed) input.

In the context of natural sciences, where calculations are performed on empirical data, the backward error is natural to consider since the perturbations on the input due to measurement imprecision might be much larger than the numerical imprecision of the computation.
In a setting where one assumes the input to be exact (up to machine precision), the forward error is arguably a more natural point of view.

Independent of whether one considers the forward or backward error, different error metrics can be used.
For example, in perturbation theory of Markov chains, $\ell_1$ and $\ell_2$ norms are considered~\cite{mitrophanov2003stability,yin2007singularly,abbas2016critical,negrea2021approximations}, 
although the maximum error on individual components is also used~\cite{yin2007singularly,abbas2016critical}. In the analysis of numerical errors of matrix multiplication algorithms, both 
errors on the individual components~\cite{miller1974computational,higham2002accuracy} and errors on the norm~\cite{demmel2007fastmatrix} have been used. A componentwise (normwise) stable algorithm is an algorithm that has a small error bound on the components (norm) of the output. 

\subsection{Numerical Stability of Matrix Multiplication}
Error analyses have been performed for various numerical algorithms, including matrix-multiplication algorithms.
The traditional $O(N^3)$ algorithm is known to be componentwise stable~\cite[Eq.(3.13)]{higham2002accuracy}, while, on the other hand, theoretically faster matrix-multiplication algorithms, such as the $O(N^{2.81})$ Strassen algorithm~\cite{strassen1969gaussian}, cannot be componentwise stable~\cite{miller1974computational} but are normwise stable~\cite{demmel2007fastmatrix}.







\section{Error Analysis}
In this section, we give bounds on the worst-case error for matrix-vector multiplication with MTBDDs, expressed in the size of the underlying system they represent. We also show how errors that approach this bound can be obtained when simulating quantum circuits. Additionally, based on these bounds, we provide suggestions for the merging threshold $\delta$ and the floating-point precision such that the error remains low.

\subsection{Framework}
\label{sec:framework}
We lay out the following framework, containing the error model and a parameterization of the problem, as a foundation not just for the error analysis when systems and their subspaces are represented as MTBDDs, but also for the future error analysis when making use of different types of decision diagrams. Furthermore, the framework applies to real and complex numbers.

\begin{enumerate}
    \item Error model:
    \begin{enumerate}
        \item We assume that the computed outcome of every numerical operation $\textsc{op}_{\comp}(\cdot,\cdot) \in \{+,-,\times,\div\}$ on two real or complex floating-point numbers yields a value equal to $\textsc{op}(a,b)(1+\theta)$ with $|\theta|\leq\varepsilon$, which is the textbook assumption on floating-point errors~\cite{higham2002accuracy}. \Cref{sec:prelims-error-analysis} provides a more detailed discussion.
        \item We assume an absolute merging threshold $\delta$, i.e. \makeleaf{} merges two values $a$ and $a'$ only if $|a - a'| \leq \delta$, thus that \makeleaf{}($a$) yields a leaf with value $a+\lambda$ with $|\lambda|\leq\delta$. A more detailed discussion is given in \Cref{sec:prelims-floats-in-dds}.
    \end{enumerate}
    \item Parameterization:
    \begin{enumerate}
        \item $n$: the size of the underlying system, expressed in the number of DD variables required to encode its states.
        \item $\varepsilon$: an upper bound on the floating-point rounding errors resulting from a single primitive operation $\{+,-,\times,\div\}$.
        \item $\delta$: an absolute threshold such that two floating-point values $a$ and $a'$ in the DD are considered equivalent if $|a-a'| \leq \delta$.
    \end{enumerate}
\end{enumerate}

\subsection{Error Bounds}
\label{sec:error-bounds}
We present componentwise forward error bounds for matrix-vector multiplication with MTBDDs, specifically for \Cref{alg:mtbdd-mult}, as a function of the size of the underlying system that generates the matrices and vectors. \Cref{thm:main_thm} gives a bound for matrices and vectors without restrictions, while \Cref{thm:main_thm-prob-quantum} gives a bound for when the input is restricted to a probabilistic or quantum system. 
\arxiv{The corresponding proofs are given in \Cref{sec:proofs}.}{The corresponding proofs are given in \cite[App. A]{arxivversion}.}
\Cref{sec:adversarial-instance} shows how the exponential terms in the error bounds can be realized. 

\begin{theorem}
\label{thm:main_thm}
   Given a weighted transition system that has states of size $n$ (giving rise to a $2^n$-sized state space and $2^n \times 2^n$ sized transition relations), the MTBDD matrix-vector multiplication algorithm that computes successor states for a transition relation (matrix) $M$ and a weighted set of states (vector) $V$ yields an output such that every entry of the resulting vector has an error of at most $(n+1) \varepsilon C + \delta2^{n+1}+O(\varepsilon^2)+O(\delta\varepsilon2^n)$, where $C=\max_{i} \sum_{j=0}^{2^n-1}|M_{i,j}V_j| \leq 2^n c_M c_V$, with $c_M$ and $c_V$ the largest absolute values of the elements of $M$ and $V$.
\end{theorem}
\Cref{thm:main_thm} can be specified to a probabilistic or quantum case, where $V$ is either a probability distribution (unit $\ell_1$-norm) and $M$ is a column stochastic matrix (preserving $\ell_1$-norm), or $V$ is a quantum state vector (unit $\ell_2$-norm) and $M$ is a unitary matrix (preserving $\ell_2$-norm).
\begin{corollary}[probabilistic/quantum case of \Cref{thm:main_thm}]
    \label{thm:main_thm-prob-quantum}
    When $M$ and $V$ are restricted to a probabilistic setting ($M$ a stochastic matrix and $V$ a probability distribution) or a quantum setting ($M$ a unitary matrix and $V$ a quantum state), every entry of the resulting vector has an error of at most $(n+1)\varepsilon +\delta2^{n+1}+O(\varepsilon^2)+O(\delta\varepsilon2^n)$.
\end{corollary}

\Cref{thm:main_thm} shows an error bound that grows exponentially in $n$, even when $\delta = 0$. As we show in \Cref{sec:adversarial-instance}, this exponential growth can be realized under adversarial assumptions.
Like \cite{demmel2007fastmatrix}, we call an algorithm numerically stable if the error grows at most polynomially. When choosing $n$ as our size parameter, we thus find that for general matrices and vectors, the MTBDD matrix-vector multiplication algorithm is not componentwise stable. 
This instability is due to the choice of size parameter. For $N \times N$ matrices, the traditional $O(N^2)$ matrix-vector multiplication is componentwise stable in $N$~\cite[Eq.(3.12)]{higham2002accuracy}, i.e. the error only grows polynomially in $N$. However, transition systems where states are of size $O(n)$ give rise to state-spaces (and thus also vectors) of size $O(2^n)$. Our analysis considers the errors to grow exponentially, not due to the algorithm itself, but due to expressing the error in terms of the underlying system size.

In probabilistic and quantum settings, MTBDD matrix-vector multiplication can be made componentwise stable in $n$ if the merging threshold $\delta$ is chosen appropriately. This is because both of these settings restrict individual values and inner products (if computed exactly) to have a magnitude of at most 1, which makes the term $C \leq 1$. This numerical stability is achieved when setting $\delta$ exponentially small in $n$. When $\delta$ is kept constant, the bound grows exponentially in $n$, as also shown in \Cref{tab:errors-for-concrete-parameters}.
The reason that merging errors (the $\delta$ terms) contribute much more than floating-point errors (the $\varepsilon$ terms) in this worst-case bound is that merging errors are additive (absolute) while floating-point errors are multiplicative (relative). As the example in the next subsection will show, when summing over exponentially many exponentially small terms, the multiplicative errors remain small when accumulated because the terms and thus also the errors are exponentially small. The size of the additive errors, however, is independent of the size of the values and will thus sum together to a term that grows linearly with the number of terms in the summation.

\begin{table}[t]
    \setlength{\tabcolsep}{0.5em}
    \centering
    \caption{Worst-case componentwise errors on the output of \Cref{alg:mtbdd-mult}, for a $2^n\times2^n$-sized unitary matrix and a $2^n$-sized vector with unit norm represented by MTBDDs, with machine precision $\varepsilon = 1.11\cdot 10^{-16}$ (smallest epsilon for 64-bit floats), merging threshold $\delta = 10^{-15}$. Without $\delta$-merging, i.e. $\delta=0$, the error 2 is zero.}
    \label{tab:errors-for-concrete-parameters}
    \begin{tabular}{c|c|c}
        $n$ & error 1: $(n+1)\varepsilon$ & error 2: $\delta2^{n+1}$\\
        \hline
        10 & $1.221\cdot 10^{-15}$& $2.048\cdot 10^{-12}$\\
        20 & $2.331\cdot 10^{-15}$& $2.097\cdot 10^{-9}$\\
        30 & $3.441\cdot 10^{-15}$& $2.147\cdot 10^{-6}$\\
        40 & $4.551\cdot 10^{-15}$& $2.199\cdot10^{-3}$\\
        50 & $5.661\cdot 10^{-15}$& $2.252\cdot10^0$\\
        60 & $6.771\cdot 10^{-15}$& $2.306\cdot10^{3}$\\
    \end{tabular}
\end{table}

\subsection{Adversarial Instance That Realizes Exponential Error}
\label{sec:adversarial-instance}
While the bounds presented in the previous section are worst-case bounds, i.e. they assume that all operations yield errors and that all errors accumulate without canceling out, we show here that the largest term in the quantum and probabilistic bounds, $\delta 2^{n+1}$, is tight, at least up to a factor 2. We also show that without the quantum or probabilistic restrictions on the input, exponential errors can be achieved even when $\delta = 0$.

Take the following $2^n \times 2^n$-sized matrix $H^{\tensor n}$ and $2^n$-sized vector $\vec x$.
\begin{align*}
    H^{\tensor n} = 
    \frac{1}{\sqrt{2}^n}
    \begin{pmatrix*}[r]
        1      & 1      & 1      & 1      & \cdots \\
        1      & \smin1 & 1      & \smin1 & \cdots \\
        1      & 1      & \smin1 & \smin1 & \cdots \\
        1      & \smin1 & \smin1 & 1      & \cdots \\
        \vdots & \vdots & \vdots & \vdots & ~\ddots
    \end{pmatrix*}
    & &  
    \vec x  = \frac{1}{\sqrt{2}^n}
    \begin{pmatrix*}[r]
        1 \\ 1 \\ 1 \\ 1 \\ \vdots
    \end{pmatrix*}
\end{align*}
The matrix $H^{\tensor n}$ represents the parallel application of $n$ of Hadamard quantum gates to $n$ quantum bits (qubits), and $\vec x$ represents a quantum state (specifically a so-called uniform superposition) made up of $n$ qubits. Both are a common occurrence in quantum computing~\cite{nielsen2010quantum}. The effect of a gate on a register of qubits can be computed through matrix-vector multiplication.

When computing $H^{\tensor n} \vec x = \vec y$, the first entry of $\vec y$ should equal
\begin{align*}
    y_0 = \sum^{2^n} \frac{1}{\sqrt{2}^n} \cdot \frac{1}{\sqrt{2}^n} = \sum^{2^n} \frac{1}{2^n} = 1.
\end{align*}
However, let us assume that the multiplication $\frac{1}{\sqrt{2}^n} \cdot \frac{1}{\sqrt{2}^n}$ yields both a floating point rounding error $|\varepsilon'| \leq \varepsilon$ and a merging error $|\delta'| \leq \delta$ (and that all additions happen without errors). The computed value will be
\begin{align*}
    \hat y_0 = 
    \sum^{2^n} \left(\frac{1}{\sqrt{2}^n} \cdot \frac{1}{\sqrt{2}^n}\right)_{\comp} = 
    \sum^{2^n} \left(\frac{1}{2^n}(1 + \varepsilon') + \delta' \right) =
    1 + \varepsilon' + \delta' 2^n,
\end{align*}
which yields an error of $|y_0 - \hat y_0| = |\varepsilon + \delta 2^n|$. The only adversarial assumption required here is that there exists some other leaf (for example from a previous computation) with value $\frac{1}{2^n}(1+\varepsilon') + \delta'$. This results in the computed value $\frac{1}{2^n}(1+\varepsilon')$ being merged with $\frac{1}{2^n}(1+\varepsilon') + \delta'$, and thus picking up a $\delta'$ error. Since both floating-point rounding and leaf merging are deterministic, all computed values pick up the same error, ultimately accumulating in an error $\delta'2^n$.

We can use a modified version of the example above to show exponentially large errors for the general setting (\Cref{thm:main_thm}), even when $\delta = 0$. This modification consists of replacing the $\sfrac{1}{\sqrt{2}^n}$ terms from $H^{\tensor n}$ and $\vec x$ with some value $a$ that is independent of $n$. By doing so, we are also not in a quantum or probabilistic setting anymore. This modification would yield an error of $|y_0 - \hat y_0| = | a^2 \cdot \varepsilon 2^{n}+ \delta 2^n|$, which grows as $O(2^n\varepsilon)$ when $\delta = 0$. An error of $O(n \varepsilon 2^n)$ can be obtained by also assuming errors on the additions.

\subsection{Parameter Suggestions}
\label{sec:delta-suggestion}
Next, we discuss suggestions for the merging threshold $\delta$ and the required floating-point precision in the probabilistic and quantum settings. These values affect both the error as well as the number of nodes in the DD. The suggested parameter values provide theoretical guarantees on the maximum error, but only serve as a heuristic for the number of DD nodes.

The bound $|y_i - \hat y_i| \leq (n+1)\varepsilon  + \delta 2^{n+1} + O(\varepsilon^2) + O(\delta\varepsilon2^n)$ can be rewritten to obtain the following suggestion for the merging threshold $\delta$, given a certain amount of $\text{allowed error} \geq (n+1)\varepsilon$.
\begin{equation}
    \delta < \frac{\text{allowed-error} - (n+1)\varepsilon}{2^{n+1}}
    \label{eq:delta-suggestion}
\end{equation}
If we want the errors on the output to grow at most polynomially with $n$, $\delta$ must thus be set as $\delta \in O(2^{-n})$. \Cref{tab:delta-suggestions} shows concrete values of $\delta$ for different errors allowed.

\begin{table}[t]
    \centering
    \caption{Values for $\delta$ such that the error is bounded by $10^{-3}$ and $10^{-6}$, assuming 64-bit floating-point values, giving $\varepsilon = 1.11 \cdot 10^{-16}$.}
    \label{tab:delta-suggestions}
    \begin{tabular}{c|c|c}
        $n$ & $\delta$ s.t. error $\leq 10^{-3}$ & $\delta$ s.t. error $\leq 10^{-6}$\\\hline
        10 & $4.883 \cdot 10^{-7}$   & $4.883 \cdot 10^{-10}$ \\
        20 & $4.768 \cdot 10^{-10}$  & $4.768 \cdot 10^{-13}$ \\
        30 & $4.657 \cdot 10^{-13}$  & $4.657 \cdot 10^{-16}$ \\
        40 & $4.547 \cdot 10^{-16}$  & $4.547 \cdot 10^{-19}$ \\
        50 & $4.441 \cdot 10^{-19}$  & $4.441 \cdot 10^{-22}$ \\
    \end{tabular}
\end{table}

However, since the purpose of $\delta$-merging is to keep the DD compact despite floating-point rounding errors, we wish that at least individual rounding errors are smaller than $\delta$. 
As mentioned in \Cref{sec:prelims-error-analysis}, floating-point values with $b$ significant bits give a relative rounding error $\varepsilon \approx 2^{-b}$. This means that for values with magnitude in the order of $2^{k}$, the absolute errors resulting from floating-point imprecision are of the order of $2^{k} \cdot 2^{-b}$ = $2^{k-b}$. For $\delta$-merging to achieve its goal, we need $\delta > 2^{k-b}$. In the probabilistic and quantum settings, where the magnitude of values never exceeds 1, we can set $k=0$, which leaves $\delta > 2^{-b}$, or 
\begin{equation}
    b > \log_2(\delta^{-1}).
    \label{eq:sig-bits-suggestion}
\end{equation}


\section{Case Study}
\label{sec:empirical-eval}
In this section, we perform a case study of the effect of different values of $\delta$ on both the numerical error and the size of the DDs during the simulation of several quantum circuits.

\subsection{Experimental Setup}
Using the decision diagram package Q-Sylvan, we run quantum circuit simulation on a selection of quantum circuits from the MQT Bench~\cite{quetschlich2023mqtbench} benchmark set.\footnote{Our setup, including instructions on how to reproduce the results, can be found online at \url{https://github.com/System-Verification-Lab/mtbdd-benchmarks}, and has been permanently archived at \url{https://doi.org/10.5281/zenodo.15322337}.}

A quantum circuit is a sequence of gates, defined by matrices, that act on an initial state, described by a vector. As a convention, the initial state for an $n$-qubit quantum circuit is the $2^n$-sized vector $\begin{pmatrix}1 & 0 & 0 & \cdots & 0\end{pmatrix}^\transpose$. If a circuit $U$ needs to run on a different input state $\vec x$, the circuit that creates $\vec x$ can simply be prepended to $U$.
For a quantum circuit with $m$ gates $U_1, \dots, U_m$, and an initial state $\vec x$, simulating the quantum circuit can be done by computing $U_m\dots U_1 \vec x$. This is typically computed as $U_m(\dots (U_1 \vec x))$, since computing $(U_m\dots U_1) \vec x$ tends to be much harder computationally, including when using decision diagrams~\cite{matsunaga1993computing}.

Since computing the effect of a single quantum gate is not a typical use case in quantum computing, and because computing the matrix that represents the whole quantum circuit is computationally infeasible, we choose to evaluate the errors when simulating the quantum circuit gate-by-gate. The corresponding pseudo-code is given in \Cref{alg:circuit-sim}.
While this yields results that show accumulated errors over multiple matrix-vector multiplications, rather than the errors on individual matrix-vector multiplications that our bounds apply to, such an evaluation is more representative of the actual use cases of DDs in quantum computing.

\begin{algorithm}[t]
\SetKwFunction{simulate}{Simulate}
\Fn{\simulate{Circuit $U$}}{
    \Comment*[l]{$U$: $n$-qubit quantum circuit with gates $U_1, \dots, U_m$}
    \BlankLine
    $S \gets \textsc{CreateInitialMTBDD}(n)$ \Comment*[r]{create MTBDD for $(1\hspace{1mm}0\hspace{1mm}0\hspace{1mm}\cdots\hspace{1mm}0)^\transpose$}
    \For{$k = 1 $ \KwTo $m$}{
        $M \gets \textsc{CreateMTBDD}(U_k)$ \Comment*[r]{create MTBDD for matrix $U_k$}
        $S \gets \textsc{Multiply}(M,S)$ \;
    }
    \Return{$S$}
}

\caption{Quantum circuit simulation using MTBDDs.}
\label{alg:circuit-sim}
\end{algorithm}

We pick three types of circuits from the MQT Bench set where MTBDDs are able to achieve good compression for the final state: the Deutsch-Jozsa (DJ) algorithm, exact quantum phase estimation (QPE), and W-state preparation. For each of these three types of circuits, the matrices that make up the circuits are categorically different, which results in different behaviors of the errors. An overview of the matrix elements for each type of circuit is given in \Cref{tab:circuits-overview}.

As a proxy for the ground truth, we use the result from a floating-point computation that uses 128 bits to represent the mantissa (with $\delta=0$) by making use of the GNU MPC library~\cite{mpclibrary}, which allows for arbitrary precision floating-point complex numbers. For the non-ground-truth computations, we use MPC values that have a 24 and 53-bit mantissa, equivalent to IEEE 754 single- and double-precision floats, which gives $\varepsilon \approx 10^{-7}$ and $\varepsilon \approx 10^{-16}$ respectively. The error metric is the largest absolute deviation from the 128-bit output, i.e. 
$
\text{the reported error} = \max_j | y^{\text{(128-bit)}}_j - \hat y_j|.
$

\begin{table}[t]
    \centering
    \def\arraystretch{1.5}
    \caption{Overview of matrix elements that occur in each type of quantum circuit.}
    \label{tab:circuits-overview}
    \begin{tabular}{c|c}
        circuit &  types of matrix elements\\\hline\hline
        Deutsch-Jozsa & $\{0, \pm 1, \pm\sfrac{1}{\sqrt{2}}, \pm \cos(\pi/4) \}$ \\\hline
        \multirow{2}{4cm}{\centering Quantum phase estimation (QPE)} & 
        \multirow{2}{7cm}{\centering $\{0, \pm 1, \pm\sfrac{1}{\sqrt{2}}, e^{i\theta}\}$, with most $\theta = -\pi/2^k$ with $k \in \{1, 2, \dots, n-2\}$ and some random $\theta$'s} \\ \\\hline
        \multirow{2}{4cm}{\centering W-state preparation} & 
        \multirow{2}{7cm}{\centering $\{0, \pm 1, \cos(\theta),\sin(\theta)\}$, for a great variety of $\theta$'s given as finite-precision decimal numbers} \\ \\\hline
    \end{tabular}
\end{table}

\subsection{Results}

We focus on the results that use a 53-bit mantissa (equivalent to 64-bit doubles), shown in \Cref{fig:qc-sim-errors-and-nodecounts}. We first discuss the results for each circuit category separately, after which we discuss general observations.

First, for the Deutsch-Jozsa circuits we find that the numerical errors are very small and mostly independent of $\delta$ and the number of qubits. This is no great surprise, as the circuits and resulting states are effectively discrete, i.e. discrete up powers of $\sfrac{1}{\sqrt{2}}$. The errors for $\delta=10^{-3}$ and $n\geq19$ are likely due to terms $(\sfrac{1}{\sqrt{2}})^n$ and $(\sfrac{1}{\sqrt{2}})^{n-1}$ becoming $\delta$-close, and thus incorrectly merge. It is interesting to note that the small errors for $\delta > 0$ are not errors at all, but rather the result of a slight imprecision in the 128-bit ground truth computation: while $\cos(\pi/4)$ should equal $\sfrac{1}{\sqrt{2}}$, the computed values differ slightly from each other. When $\delta = 0$, as it is for the ground-truth computation, occurrences of $\sfrac{1}{\sqrt{2}}$ and $-\cos(\pi/4)$ fail to perfectly cancel each other out, leaving a result with several very small values that should instead equal 0. Setting $\delta > 0$ allows for the occurrences of $\cos(\pi/4)$ to merge with $\sfrac{1}{\sqrt{2}}$ and very small values to be merged with 0. This shows that for such ``almost discrete'' instances, having a small non-zero $\delta$ not only benefits the DD size, but can also correct small errors in the computation. 

Next, for the QPE circuits, the $e^{i\theta}$ terms complicate the computation in two ways: the random $\theta$'s increase the number of unique values that do not have an exact floating-point representation, and $\theta$'s of the form $\theta = -\pi/2^k$ introduce values that are exponentially small in $n$. When either $\delta$ or $n$ becomes too big, the exponentially small values are merged with 0, which causes significant errors in the result. The coincidence between erroneous instances and large DDs is caused by the errors destroying the symmetry that allows the DDs to stay compact.

{
\newcommand{\scale}{0.43}
\begin{figure}[p]
    \centering
    \captionsetup[subfloat]{farskip=3mm,captionskip=0mm}
    \subfloat{%
        \includegraphics[scale=\scale]{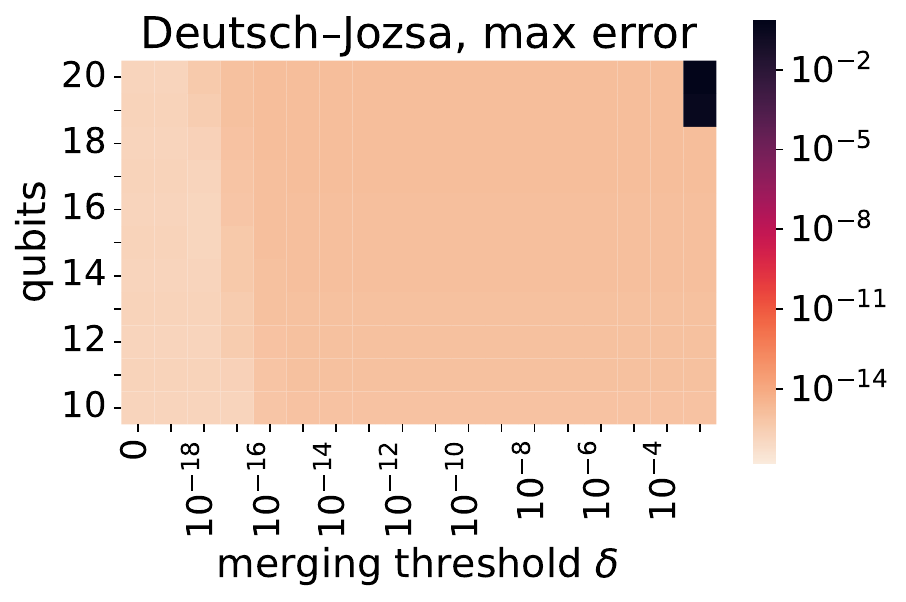}%
        \label{fig:dj-max-abs-error}%
    }
    \subfloat{%
        \includegraphics[scale=\scale]{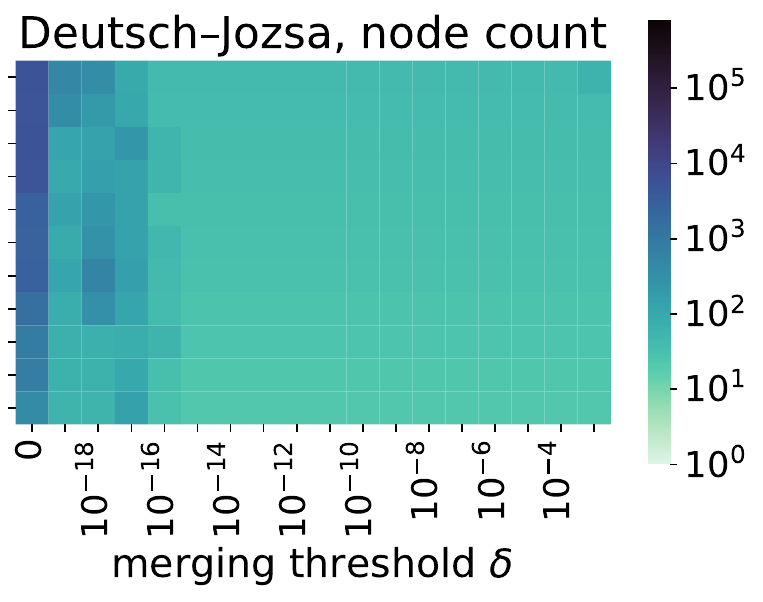}%
        \label{fig:dj-max-nodes}%
    }
    \\
    \subfloat{%
        \includegraphics[scale=\scale]{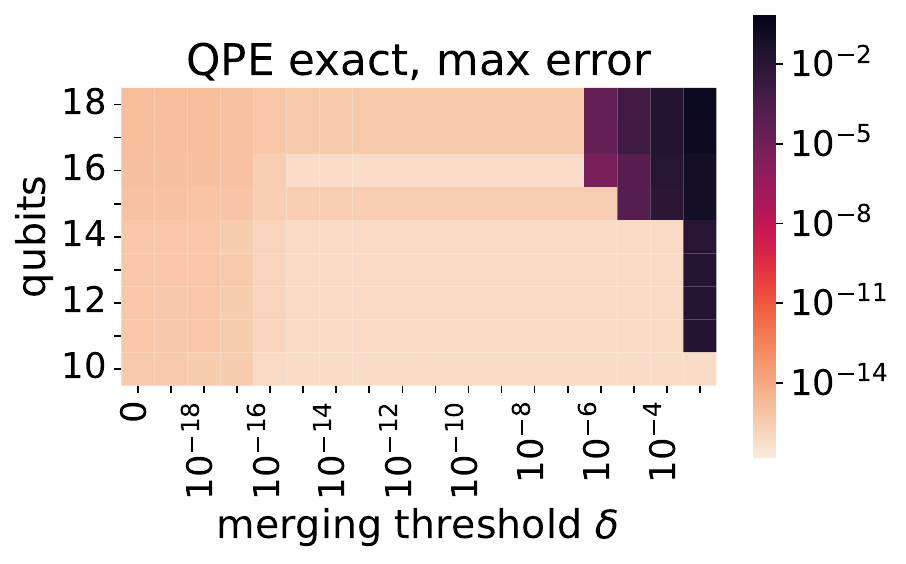}%
        \label{fig:qpeexact-abs-error}%
    }
    \subfloat{%
        \includegraphics[scale=\scale]{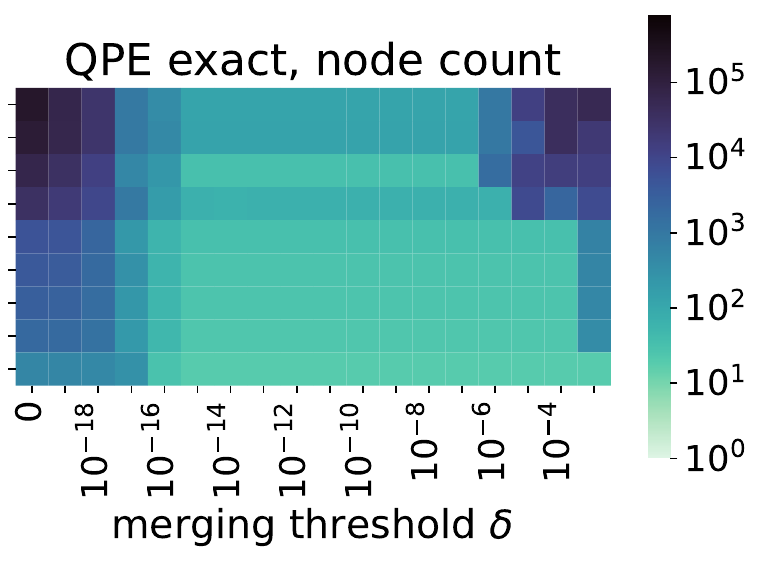}%
        \label{fig:qpeexact-max-nodes}%
    }\\
    \subfloat{%
        \includegraphics[scale=\scale]{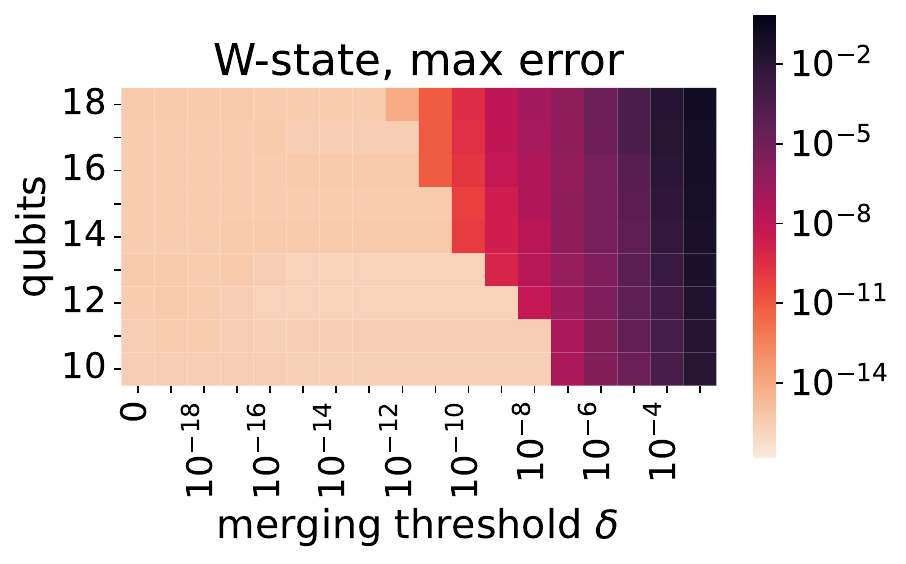}%
        \label{fig:wstate-abs-error}%
    }
    \subfloat{%
        \includegraphics[scale=\scale]{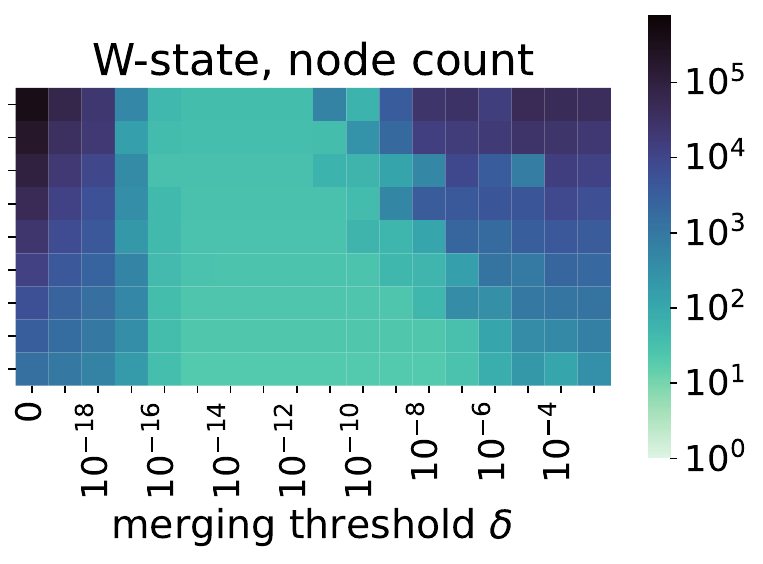}%
        \label{fig:wstate-max-nodes}%
    }
    \caption{Errors and number of DD nodes of the final state for different circuits, with $b=53$ mantissa bits, equivalent to double precision floats.
    }
    \label{fig:qc-sim-errors-and-nodecounts}
\end{figure}
}

{
\newcommand{\scale}{0.43}
\begin{figure}[p]
    \centering
    \captionsetup[subfloat]{farskip=3mm,captionskip=0mm}
    \subfloat{%
        \includegraphics[scale=\scale]{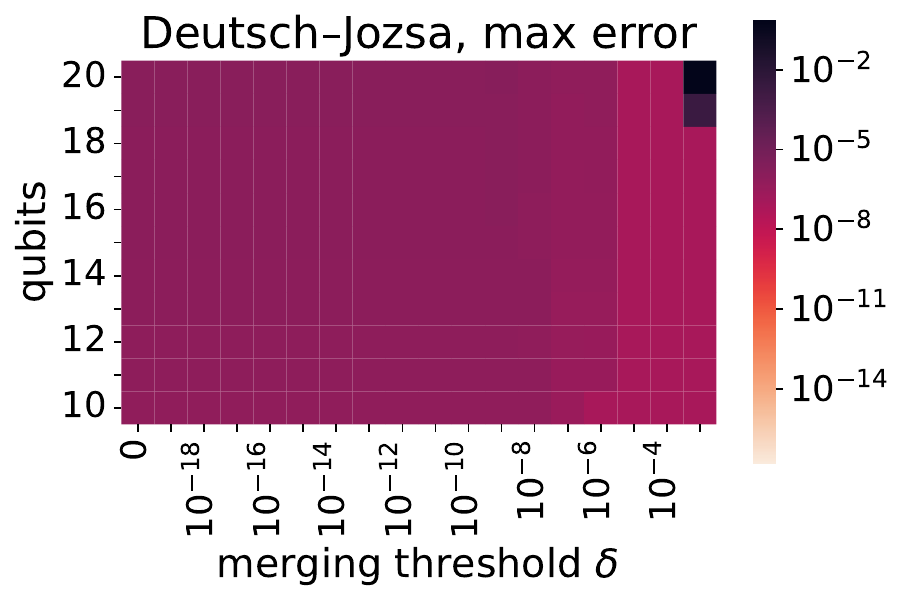}%
        \label{fig:dj-max-abs-error24}%
    }
    \subfloat{%
        \includegraphics[scale=\scale]{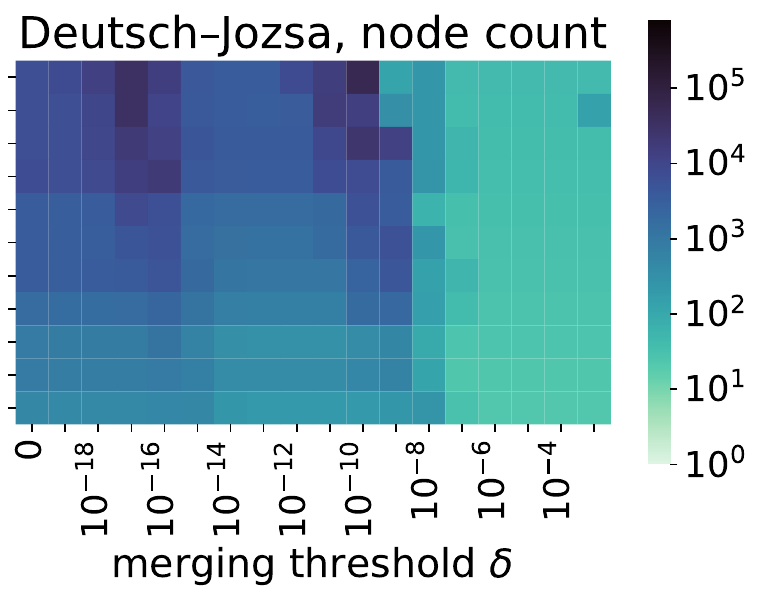}%
        \label{fig:dj-max-nodes24}%
    }
    \\
    \subfloat{%
        \includegraphics[scale=\scale]{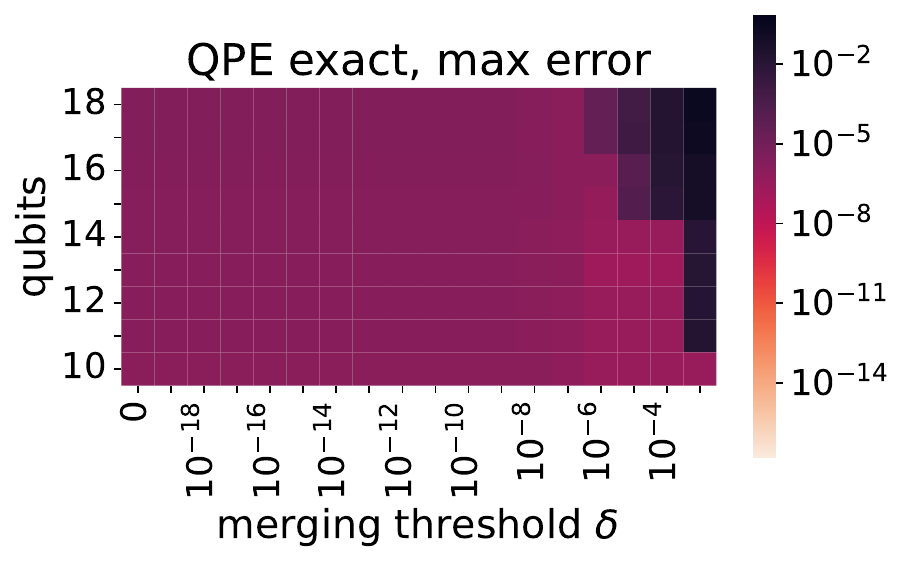}%
        \label{fig:qpeexact-abs-error24}%
    }
    \subfloat{%
        \includegraphics[scale=\scale]{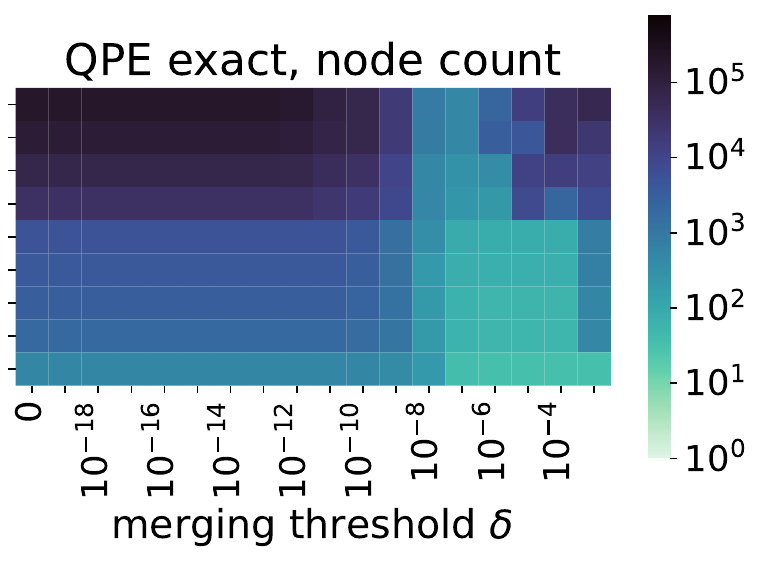}%
        \label{fig:qpeexact-max-nodes24}%
    }\\
    \subfloat{%
        \includegraphics[scale=\scale]{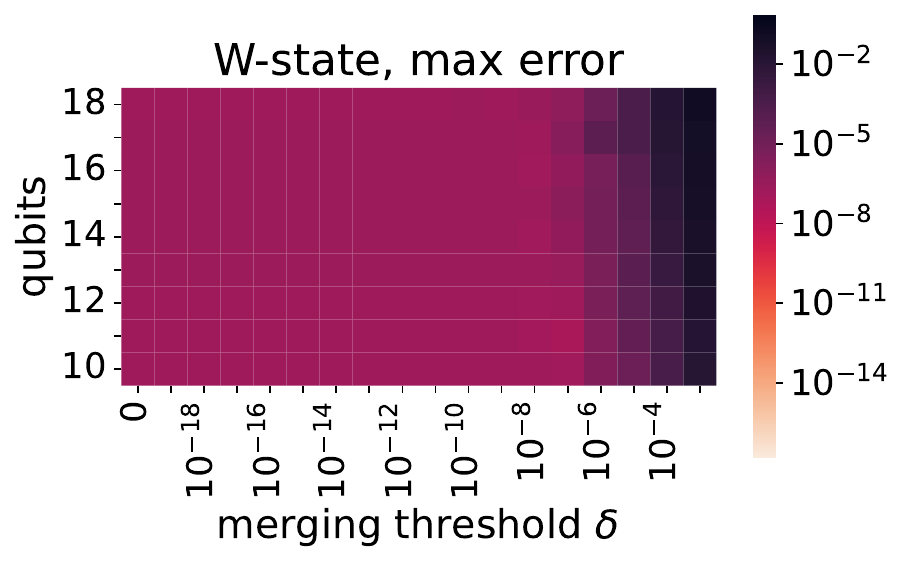}%
        \label{fig:wstate-abs-error24}%
    }
    \subfloat{%
        \includegraphics[scale=\scale]{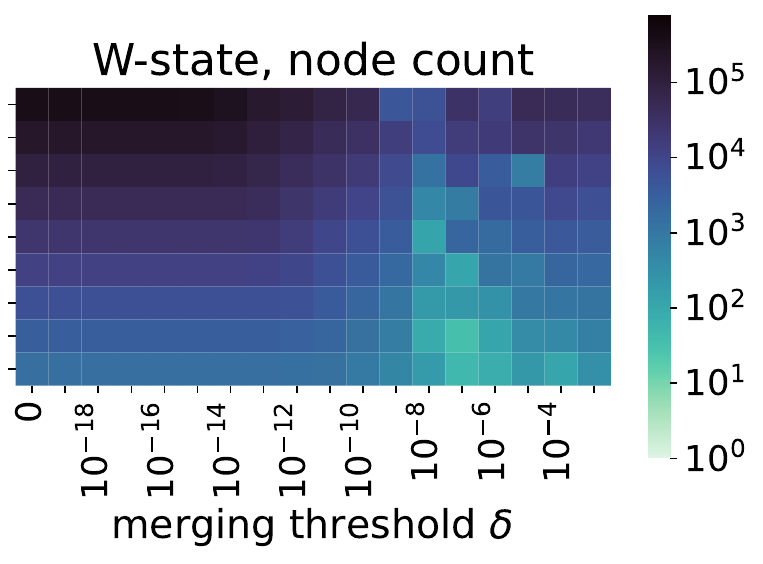}%
        \label{fig:wstate-max-nodes24}%
    }
    \caption{Errors and number of DD nodes of the final state for different circuits, with $b=24$ mantissa bits, equivalent to single precision floats.
    }
    \label{fig:qc-sim-errors-and-nodecounts-24bits}
\end{figure}
}

Finally, the W-state circuits contain many different values that do not have exact floating-point representations. Under an exact computation, these values should all multiply and sum together such that the resulting vector only contains values in $\{0,\sfrac{1}{\sqrt{n}}\}$. However, the complexity with which this happens leaves a lot of opportunity for errors to be introduced. 
As $\delta$ and $n$ increase, the errors for the W-state circuit increase more gradually than for the QPE circuit. This is likely due to the W-state circuit producing more unique values during the computation, resulting in merging errors occurring more frequently but affecting fewer values when they do.

For all three circuits, we find that when $\delta = 0$ the errors are small (as also discussed in \Cref{sec:error-bounds}), but the number of DD nodes is also significantly greater than it should be if the calculation was exact. As discussed in \Cref{sec:prelims-floats-in-dds}, this is caused by sub-functions that should theoretically be equal no longer being recognized as such due to small floating-point errors. This illustrates that in order to benefit from the compression DDs provide, some sort of $\delta$-merging must be used when working with floating-point values.

The issue with $\sfrac{1}{\sqrt{2}}$ and $\cos(\pi/4)$ not being recognized as equal also highlights that the exact choice of gates matters. For example, it is likely that decomposing a Toffoli gate (a 3-qubit gate made up of 1's and 0's) into single- and two-qubit gates (which contain terms $\sfrac{1}{\sqrt{2}}$ and $e^{i\pi/4}$) before simulation yields significantly worse numerical errors.

\Cref{fig:qc-sim-errors-and-nodecounts-24bits} shows errors and DD sizes for MPC values that use a 24-bit mantissa (equivalent to 32-bit floats). It is unsurprising that a lower precision results in greater errors. However, considering the node counts in \Cref{fig:qc-sim-errors-and-nodecounts,fig:qc-sim-errors-and-nodecounts-24bits}, we find that the range of $\delta$ that yields compact DDs gets wider (narrower) when the number of mantissa bits increases (decreases). Additionally, the node counts for the QPE and W-state circuits show that this range decreases as $n$ increases. This advocates for DD implementations that can increase the precision of the floating-point numbers as $n$ increases, to maximize their compactness. 


\section{Conclusion}
\label{sec:conclusion}
\paragraph{Summary}
In this paper, we have analyzed the numerical stability of the matrix-vector multiplication algorithm for MTBDDs, which is a crucial component in the exhaustive exploration of discrete, probabilistic, and quantum systems. We have found that the numerical errors that can accumulate during the execution of this algorithm can be kept polynomially small relative to the size $n$ of the system under investigation, only if certain parameters are appropriately restricted. Specifically, for quantum and probabilistic systems, to keep the errors from growing exponentially in $n$, the merging threshold should be set exponentially small in $n$.

In a case study on a selection of quantum circuits, we have shown that while adversarial instances can realize exponentially large errors, the practical effect of numerical errors can greatly vary between different types of instances. The empirical results also suggest that at least in quantum use cases, a good heuristic for setting the merging threshold $\delta$ is to set it as small as possible while keeping $\delta > 2^{-b}$, where $b$ is the number of bits in the mantissa of the floating-point representation. Ideally, $b$ can grow with the system size $n$, however, when restricted to 64-bit doubles, setting $\delta$ to $10^{-15}$ appears to be the best trade-off between errors and DD size.


\paragraph{Future Work}
For our error bounds, we have picked the size of (the states of) the underlying system as the size parameter, however, another natural size parameter for the analysis of DD algorithms is the number of DD nodes.
While our error analysis is completely independent of the number of nodes, it would be interesting to investigate if a different analysis can be made, either for MTBDDs or other DDs, that relates the error and the number of DD nodes.

While our results show both theoretical and empirical insights into the numerical errors in computations done with MTBDDs, other types of DDs exist that can offer more compactness than MTBDDs. These include EVDDs~\cite{miller2006qmdd,zulehner2018advanced}, CFLOBDDs~\cite{sistla2023symbolic}, and LIMDDs~\cite{vinkhuijzen2023limdd,hong2025limtdd}.
Our error analysis and the characterization of the problem in terms of the number of variables $n$, the floating-point errors $\varepsilon$, and the merging threshold $\delta$ can serve as a foundation for the analysis of other DD algorithms as well as other types of DDs.


\bibliographystyle{splncs04}
\bibliography{references} 

\arxiv{
\appendix
\section{Proofs}
\label{sec:proofs}

In this section, we formally analyze the forward numerical error of matrix-vector multiplication with MTBDDs and thus prove \Cref{thm:main_thm} and \Cref{thm:main_thm-prob-quantum}.

\subsection{Error model and notation}
As discussed in \Cref{sec:prelims-error-analysis}, we assume that every numerical operation $\textsc{op}(\cdot,\cdot)$ on two real or complex numbers yields a computed value equal to $\textsc{op}(a,b)(1+\theta)$ with $|\theta|\leq\varepsilon$.
For the \makeleaf{} algorithm, we assume values are merged based on their absolute difference, as discussed in \Cref{sec:prelims-floats-in-dds}. This means $\makeleaf(a)$ results in a leaf with value $a+\lambda$ with $|\lambda|\leq\delta$, where merging parameter $\delta$ can be freely chosen. 
We define $\Theta=\{1+\theta\mid|\theta|\leq\varepsilon\}$ and $\Lambda=\{\lambda\mid|\lambda|\leq\delta\}$. For two sets $A,B$ we define $A+B = \{a+b\mid a\in A, b\in B\}$ and $A\cdot B = \{a\cdot b\mid a\in A, b\in B\}$. In particular, we have $\Theta^j=\underbrace{\Theta\cdot\Theta\cdot\dots\cdot\Theta}_{j \text{ elements}}$.

\subsection{Analysis of matrix-vector multiplication algorithm for MTBDD}
\label{sec:mat-vec-mult-analysis}

We analyze the multiplication of an $2^n\times 2^n$ sized matrix $M$ with an $2^n$ sized vector $V$, both represented by an MTBDD. The algorithm is stated in \Cref{alg:mtbdd-mult}.

As we analyze the numerical stability, we only take care of operations on numerical values, which are addition and multiplication of two numerical values and rounding in the \makeleaf{} algorithm. We do a worst-case analysis, so we assume that all errors accumulate and never cancel out against each other.

For the analysis of the numerical errors the caching of \mult and \add operations can be ignored, as the numerical values that are obtained are equivalent with and without caching. A result from cache contains equivalent numerical errors as calculating the result again. \\
To be complete, one should note that caching causes a correlation of numerical errors: the numerical errors of a cache-entry can be used multiple times. This does not influence our analysis.

Now, we start the analysis of the algorithm. When following the recursion on the call of \mult{$M,V$}, we see that first multiplications of values $M_{i,j}$ and $V_j$ are applied. This results in a value $(M_{i,j}V_j)\cdot\Theta$. Then a \makeleaf{} is applied, so the computed value of the leaf is then $(M_{i,j}V_j)_{\comp} = (M_{i,j}V_j)\cdot\Theta+\Lambda$. Then the \add algorithm recursively adds these values for all $j$.

The binary recursion is as follows:
\begin{align*}
    &\left(\sum_{j=0}^{2^{m+1}-1}M_{i,j}V_j\right)_{\comp} \\
    &\qquad=\left[\left(\sum_{j=0}^{2^{m}-1}M_{i,j}V_j\right)_{\comp} + 
    \left(\sum_{j=2^m}^{2^{m+1}-1}M_{i,j}V_j\right)_{\comp}\right]\Theta+\Lambda.
\end{align*}
The $\Theta$ comes in because the addition of two values introduces an error in $\Theta$. The $\Lambda$ comes in because every time two values are added they are rounded by the \makeleaf{} algorithm.

Note that the recursion is slightly more general, as there are other branches in the recursion. It turns out that all branches behave similarly: for $k \in \{0$, $2$, $4$, $\dots$, $2^{n-m}-2\}$
\begin{align*}
    &\left(\sum_{j=k2^{m}}^{(k+2)2^{m}-1}M_{i,j}V_j\right)_{\comp} \\
    &\qquad= \left[\left(\sum_{j=k2^m}^{(k+1)2^{m}-1}M_{i,j}V_j\right)_{\comp} + 
    \left(\sum_{j=(k+1)2^m}^{(k+2)2^{m}-1}M_{i,j}V_j\right)_{\comp}\right]\Theta+\Lambda.
\end{align*}

\begin{figure}[t]
    \centering
    \scalebox{0.97}{$
    \underbrace{
        \underbrace{
            \underbrace{(M_{i,0}V_0)_{\comp}+(M_{i,1}V_1)_{\comp}}_{\left(\sum_{j=0}^1 M_{i,j}V_j\right)_{\comp}} + 
            \underbrace{(M_{i,2}V_2)_{\comp}+(M_{i,3}V_3)_{\comp}}_{\left(\sum_{j=2}^3 M_{i,j}V_j\right)_{\comp}}
        }_{\left(\sum_{j=0}^3 M_{i,j}V_j\right)_{\comp}} + 
        \dots + 
        (M_{i,2^n-1}V_{2^n-1})_{\comp}
    }_{\left(\sum_{j=0}^{2^n-1} M_{i,j}V_j\right)_{\comp}}
    $}
    \caption{Visualization of the recursive summation as applied in the matrix-vector multiplication algorithm with MTBDDs.}
    \label{fig:binary_recursive_formula}
\end{figure}
Working out this recursion leads to the formula
\begin{equation}
    \left(\sum_{j=0}^{2^{n}-1}M_{i,j}V_j\right)_{\comp} = \sum_{j=0}^{2^n-1}M_{i,j}V_j\Theta^{n+1} + \sum_{j=0}^n2^j\Lambda\Theta^j.
\end{equation}

This value is the value of the $i$'th entry of the multiplication result and is stored in one of the leaves of the MTBDD representing the output vector. 
Thus, the difference between the computed value and the exact value of a leaf node, representing a vector entry of the final vector, is
\begin{align*}
    &\left|\left(\sum_{j=0}^{2^{n}-1}M_{i,j}V_j\right)_{\comp} - \sum_{j=0}^{2^{n}-1}M_{i,j}V_j\right| \\
    &\qquad\leq \sum_{j=0}^{2^{n}-1}\left|M_{i,j}V_j\right|\cdot\left|\Theta^{n+1}-1\right|+\sum_{j=0}^n 2^j|\Lambda|\cdot|\Theta^j| \\
    &\qquad\leq ((1+\varepsilon)^{n+1}-1)\sum_{j=0}^{2^n-1}|M_{i,j}V_j| + \delta\sum_{j=0}^n2^j(1+\varepsilon)^j
\end{align*}

The last equality holds because every value in $\Theta$ can be bound by $1+\varepsilon$ and every value in $\Lambda$ can be bound by merging parameter $\delta$. 

Thus we have for every computed value $\hat y_i$ in the leaves of the resulting MTBDD when multiplying matrix $M$ with vector $V$:
\begin{align}
    |y_i- \hat y_i| &\leq ((1+\varepsilon)^{n+1}-1)\sum_{j=0}^{2^n-1}|M_{i,j}V_j| + \delta\sum_{j=0}^n2^j(1+\varepsilon)^j\\
    \label{eq:MTBDD-error-general-case}
    &= (n+1)\varepsilon C + \delta2^n+ O(\varepsilon^2) + O(\delta\epsilon2^n)\\
    & \qquad\qquad\text{for }  C=\max_{i}\left[\sum_{j=0}^{2^n-1}|M_{i,j}V_j|\right]\leq2^n\max_{i,j}|M_{i,j}|\max_j|V_j|.
\end{align}
This proves \Cref{thm:main_thm}.\qed

Now we will prove \Cref{thm:main_thm-prob-quantum} for quantum systems.\\
When we assume that the matrix-vector multiplication is a quantum gate multiplied with a quantum state, we have that $M$ is a unitary matrix and $V$ is a quantum state, so it has $\ell_2$-norm equal to 1. Therefore we can write
\begin{align*}
    C=\max_i\left[\sum_{j=0}^{2^n-1}|M_{i,j}V_j| \right]
    &\leq \max_{j}\{|V_{j}|\}\max_i\left[\sum_{j=0}^{2^n-1}|M_{i,j}|\right] \\
    &\leq \max_{j}\{|V_{j}|\}\max_i\left[\left(\sum_{j=0}^{2^n-1}|M_{i,j}|^2\right)^{1/2} \right] \\
    &=\max_{j}\{|V_{j}|\}
    \leq1.
\end{align*}
The second inequality is due to the Cauchy-Schwarz inequality. The last equality holds because $M$ is a unitary matrix. The last inequality holds because $V$ has $\ell_2$-norm 1.
Thus for quantum gate updates of states we can rewrite \eqref{eq:MTBDD-error-general-case} as:
\begin{align}
    |y_i - \hat y_i| &\leq (n+1)\varepsilon + \delta2^n+ O(\varepsilon^2) + O(\delta\epsilon2^n).
\end{align}

Now we will prove \Cref{thm:main_thm-prob-quantum} for probabilistic systems.\\
We assume that the matrix-vector multiplication is a stochastic matrix multiplied with a probability distribution vector. Therefore we can write
\begin{align*}
    C=\max_i\left[\sum_{j=0}^{2^n-1}|M_{i,j}V_j|\right]\leq \max_{i,j}\{|M_{i,j}|\}\sum_{j=0}^{2^n-1}|V_{j}|=\max_{i,j}\{|M_{i,j}|\}\leq1.
\end{align*}
The last equality holds because $V$ is a probability vector matrix. The last inequality holds because $M$ is a stochastic matrix.
Thus for Markov chain updates of a probability distribution we can rewrite \eqref{eq:MTBDD-error-general-case} as:
\begin{align}
    |y_i - \hat y_i| &\leq (n+1)\varepsilon + \delta2^n+ O(\varepsilon^2) + O(\delta\epsilon2^n).
\end{align}
This proves \Cref{thm:main_thm-prob-quantum}.\qed


}{}

\end{document}